\documentclass{article}

\usepackage[english]{babel}
\usepackage{authblk}

\usepackage[letterpaper,top=2cm,bottom=2cm,left=3cm,right=3cm,marginparwidth=1.75cm]{geometry}

\usepackage{amsmath}
\usepackage{graphicx}
\usepackage[colorlinks=true, allcolors=blue]{hyperref}

\begin{document}

\title{Overprocurement of balancing capacity may increase the welfare in the cross-zonal energy-reserve coallocation problem}
\author[1,2,*]{D\'{a}vid Csercsik}
\author[3]{\'{A}d\'{a}m Sleisz}

\affil[1]{\small Institute of Economics, ELTE Centre for Economic and Regional Studies T\'oth K\'alm\'an u. 4., H-1097 Budapest, Hungary\\ \emph{csercsik.david@krtk.elte.hu}}
\affil[2]{\small P\'{a}zm\'{a}ny P\'{e}ter Catholic University, Faculty of Information Technology and Bionics, Pr\'{a}ter~u. 50/A H-1083 Budapest, Hungary}
\affil[3]{\small Budapest University of Technology and Economics, Department of Electric Power Engineering\\ Egry J. u. 18. 1111 Budapest, Hungary \\
              \emph{sleisz.adam@bme.hu}}
\affil[*]{Corresponding author}

\maketitle

\begin{abstract}
When the traded energy and reserve products between zones are co-allocated to optimize the infrastructure usage, both deterministic and stochastic flows have to be accounted for on interconnector lines. We focus on allocation models, which guarantee deliverability in the context of the portfolio bidding European day-ahead market framework, assuming a flow-based description of network constraints. In such models, as each unit of allocated reserve supply implies additional cost, it is straightforward to assume that the amount of allocated reserve is equal to the accepted reserve demand quantity. However, as it is illustrated by the proposed work, overprocurement of reserves may imply counterintuitive benefits. Reserve supplies not used for balancing may be used for congestion management, thus allowing valuable additional flows in the network.
\end{abstract}

\section{Introduction}

Since the trade of products related to electrical energy requires designated infrastructure, transmission constraints of the underlying network must always be considered. Accordingly, these products are traded within the frameworks of various dedicated markets, which coordinate the respective transactions and estimate network states in advance to avoid congestion and operational hazards.
Trading of different products corresponds to different types of network loads.
Transactions of energy products scheduled in advance implies deterministic flows in the network. Various wholesale electricity market design approaches take the energy flows implied by the transactions into account on different levels. While nodal (or locational marginal) pricing models use a detailed description of the network \cite{liu2009locational}, zonal pricing approaches define groups of nodes (i.e., zones) and assume congestion may occur only between zones \cite{weibelzahl2017nodal}. In the current paper, we follow the terminology and assumptions of the latter approach and assume zones, inside which no potentially congested elements are present.

The trading of electrical energy products aims to match the predicted consumption with the planned production. However, the demand of certain consumers (e.g., domestic) and the production levels of some generating units (e.g., weather-dependent renewable sources) can not be perfectly predicted.
Nevertheless, any imbalance between the total consumed and produced power results in frequency deviation, which can jeopardize the system stability. To maintain the equality of production and consumption, and reduce the imbalance to zero, ancillary services and products for network operation management are in use.
These services are required and bought by transmission system operators (TSOs), who are responsible for the operational stability of the power grid.
Ancillary services, or in other terms, reserve products, are fundamentally options for energy consumption or production defined on various time scales \cite{frunt2010classification} that may be activated when necessary, to deal with imbalances.
For the sake of simplicity, in this paper, we describe reserve bids and allocations 
using their underlying energy quantity instead of power values. We assume
hourly reserve procurement; therefore, a reserve bid of $q$ MWs is indicated as its $q$ MWh underlying energy.
We distinguish between positive and negative reserve products. Activating a positive reserve product corresponds to the ramping up of a controllable generating unit (or ramping down of a controllable load), while activating a negative reserve product corresponds to the ramping down of a controllable generating unit (or ramping up of a controllable load).
% few words about coordinated procurement..

From the perspective of the network operator, the best scenario is when the balancing energy is activated in the same zone where the imbalance arises. This way, the balancing activity does not put additional load on critical interzonal network elements representing bottlenecks of the system. Due to various reasons, however, this is not always necessary or desired in other contexts. First, this may be technically impossible. Balancing may be delivered only by controllable generators, controllable loads, or fast-response storage units (that can act as both, depending on actual load level), which are not necessarily present at every zone in the network, or their capacity may not meet the reserve requirements in the actual area. Furthermore, even if such units are present, it may be economically more profitable to allocate reserve resources outside the zone to sources that can provide the necessary services at a lower price. Regarding the price of ancillary products, two cost components can be distinguished. The allocation cost is paid for the provider of the product to be on a stand-by and keep the unit prepared to implement the defined operational change (up- or down-ramping) of the unit when necessary, while the activation cost arises only if the reserve product is activated, and it is typically proportional to the rate of activation.

\subsection{Coordination of reserve activation}
%Some words about reserve demand netting...
The article \cite{van2018cross} defines two basic paradigms of reserve activation: (i) common merit order and (ii) imbalance netting. 
While the common merit order means that, independent of their location, always the reserve resources with the lowest possible activation cost are activated to match the actually arising reserve demand, imbalance netting is based on the principle that up and down reserve demands in different zones can be netted, reducing the amount of reserve sources that must be activated in the final dispatch.
Both activation paradigms require the availability of network transmission capacities, since inter-zonal reserve activation induces power flows in the network. If the network capacity available for these reserve-activation induced flows is limited, the set of activated resources must be accordingly adjusted (e.g. not the cheapest set of resource providing units will be activated, but the cheapest \emph{deliverable } configuration will be determined).  
There are already operating platforms for different reserve products in the European Union.
The MARI (Manually Activated Reserves Initiative) is a coordination platform for manually activated frequency control reserves (mFRR) \cite{roumkos2021manual}, while the acronym PICASSO stands for the Platform for the International Coordination of Automated Frequency Restoration and Stable System Operation \cite{backer2023economic}. In addition, the TERRE project (Trans European Replacement Reserve Exchange) aims to coordinate the exchange of replacement reserves \cite{caprabianca2020replacement}.

%Imbalance netting is based on the consideration that up and down reserve demands between different zones may be netted before the activation process of reserve resources.

\subsection{Coordination of reserve allocation and deliverability}

To ensure the safe operation of the power grid, ancillary services must be procured and allocated in advance. 
The inter-zonal (or cross-zonal) allocation of reserves, however, implies stochastic flows in the network calculations, since at the time of the allocation, it is uncertain whether the allocated reserve will be activated or not. The allocation of reserves must be carried out in a way that ensures the later deliverability of the reserves, i.e., for any possible realization of the reserve demand, a reserve supply activation pattern must exist, for which the implied flows are operationally feasible.
What are the conditions of such a feasibility?

If one considers the market-based procurement and allocation of reserve products, it is necessary to notice that energy and reserve markets are interdependent
since they make use of the same generation and consumption capabilities
of market actors, and they both utilize the transfer capacity of the power grid.
Because of this interdependence, it is a generally accepted viewpoint that the coordinated trading and market clearing of
energy and reserve products is desirable from an economic perspective \cite{gonzalez2014joint}. Some authors also argue that the benefits of such a co-allocation are expected to grow with the increasing share of renewable and distributed generation \cite{van2020energy}.

\subsection{Motivation}

In this subsection, we briefly review the models proposed so far that aim to address the issue of reserve deliverability.
The first group of results corresponds to articles that do not clear the energy and the reserve markets simultaneously, i.e., no co-allocation of energy and reserve products is executed.
%[11]
The article \cite{van2018cross} assumes a sequential clearing of the energy and reserve markets. This approach implies that the transactions of the energy market have priority regarding network capacity allocation, and only the remaining capacity of the network may be used for the trading of reserve products. The results derived in this article are based on a unit commitment model that includes no explicit parameter describing the heterogeneity of reserve allocation costs over generation units. In addition, the work described in \cite{van2018cross} uses a simplified network representation, i.e., no flow-based network model is applied. 
%[12]
The approach proposed in \cite{gebrekiros2015reserve} may be regarded in some sense as the dual of \cite{van2018cross}, as in this model a successive scheme is applied, where frequency restoration reserves are allocated first, followed by the clearing of the day-ahead market, taking into account the transmission capacity allocated for possible future cross-border reserve activation. The network description is similar to \cite{van2018cross}, an NTC (network transfer capacity) based approach is used to account for flows of energy trade and inter-zonal reserve activation.
The model proposed in \cite{gebrekiros2015reserve}, however, already includes the reserve bidding prices of various units, representing the cost heterogeneity of reserve allocation.
%[16]
The article \cite{karaca2022enabling} represents a significant advancement compared to the above detailed approaches, as it applies a preemptive transmission allocation model, which aims to optimally divide the transmission capacity between inter-zonal energy and reserve trading before the clearing of the respective markets. It also includes reserve bidding prices and uses a flow-based network representation. Nevertheless, the energy and reserve markets are also cleared sequentially in this approach as well.
%[17]
% \cite{viafora2020dynamic} - ez a koallokacio hataran van (bilevel prog), de a zonak meghatarozasara fokuszal inkabb

The second group of results corresponds to papers, where energy and reserves are allocated in a single step (co-allocation).
%[13]
The paper \cite{ihlemann2022benefits} proposes a model for the joint clearing of day-ahead energy and balancing capacity markets, with coordinated sizing and procurement of balancing capacity.
The available cross-zonal transmission capacity in this model is optimally allocated for those energy/reserve flows that imply the most cost-saving. This paper also uses a unit-commitment type description, in which the demand elasticity of balancing capacity is assumed to be zero. In addition, no flow-based network representation or reserve allocation bids are included in the model. 
% [14] Chen
The model described in \cite{chen2013incorporating} also assumes the simultaneous allocation of energy and reserves, ensuring deliverability by incorporating post reserve deployment transmission constraints. While this model also derives zonal market-clearing
prices (MCPs) for reserve, it does not include the description of the heterogeneity of reserve allocation costs or reserve bidding prices.
%Let us furthermore mention the papers 
%\cite{viafora2020dynamic} and \cite{wang2014dynamic}, proposing the concept of dynamical reserve zone.
%[15] Wang
\cite{wang2014dynamic} proposes a model in which the zones are not predefined, but are determined by a clustering algorithm. Energy dispatch and reserve allocation are jointly determined by a unit-commitment algorithm.

The above results are dominantly based on the unit-commitment modelling approach. Unit-commitment models are based on the technical-economic parameters of generating units and assume that these units are not strategic market players. In addition, these models do usually assume fixed demand and do not account for demand flexibility, thus they have limited potential to describe European day-ahead markets, where the central concept is on self-scheduling and marginal prices without uplift payments.

The recent paper \cite{divenyi2024potential} analyzes the problem outside the scope of unit-commitment models. The article, which is based on the portfolio-bidding formalism used on the European day-ahead market, proposes a novel energy-reserve co-allocation scheme based on robust optimization that ensures the deliverability of the allocated reserves, assuming a flow-based network representation, and also handles nonconvex orders, which are essential characteristic features of day-ahead power markets. 

We argue that the optimal cross-zonal reserve allocation problem is so complex that there are still some aspects that are not captured, even by this recent advanced approach.
While the usual assumption of unit-commitment approaches is that the available reserve must be at least equal to the predefined demand, the co-allocation framework of \cite{divenyi2024potential} assumes that the allocated reserve (the total quantity of accepted reserve supply offers) equals the actual reserve demand (total quantity of accepted reserve demand bids).
In the used portfolio-bidding framework, the assumption seems straightforward at first glance. Both for positive and negative reserves, the allocated supply must be at least equal to the demand, to allow TSOs to satisfy reserve demands also in the extreme case, when one type (up or down, i.e., positive or negative) must be fully activated and no activation request is present for the other type, thus no imbalance netting is possible. 
On the other hand, in the portfolio-bidding framework, each unit of allocated reserve implies additional cost, so it seems sound to ask, why allocate more reserve supply compared to the demand? This second consideration will be the subject of this manuscript.

\subsection{Contribution}
In this paper, we illustrate one counterintuitive aspect of the optimal cross-zonal reserve allocation problem.
We show that the overprocurement of reserves may result in additional flexibility during the activation process, more precisely, allocated reserves not used for
balancing may be used for congestion-alleviation, and the benefit of additional transfer capacity can overwhelm the extra costs of allocating extra units or reserve over the demand quantity. Let us note that if both positive and negative reserves are allocated in a network, some reserves will be unused in the activation, since, in the context of individual zones, it is not possible that positive and negative reserve demand arises at the same time (as in this case, it could be netted). We will see, however, that unused reserves can be used for congestion management only, if free positive and negative reserve resources are simultaneously present in the system after the balancing needs have been satisfied. In the case of extreme activation (maximal up or down) this is only
possible, if at least one reserve type is over-allocated.

Context of the analysis: Neglecting activation costs. Similar to \cite{divenyi2024potential}, we assume that the activation costs are not known at the time of allocation.

%In essence, the separation of the procurement and activation steps is
%feasible due to the guarantee itself. This separation is realistic because
%the exact price of activation is generally not known when procurement
%happens, and therefore no reliable economic reasoning exists to influence
%the activation process. On the other hand, the separation is also
%practical because there are already operating independent reserve
%activation platforms in Europe (MARI [9] for manual and PICASSO [10]
%for automatic activation). Furthermore, the imbalances of different European
%transmission system operators (TSOs) are frequently netted in
%practice to resolve them without reserve activation.

\section{Methodology}
We use a very simple energy-reserve coallocation model.
Three products are considered in the proposed model: Energy (E), Positive, and negative reserve (R+ and R-, respectively).
Market participants submit simple divisible bids described by quantity-price pairs for each product in each zone. Quantities for supply bids are taken into account as negative. The network flows implied by interzonal transfers are calculated based on power transfer distribution factors (PTDFs) \cite{vsovsic2014features}.

We require the clearing algorithm to determine the set of accepted/rejected (or partially accepted) bids and the market-clearing prices for each traded product for each zone (we consider a single period), considering the required balance of product over all zones,
The bid acceptance indicators have to match the market-clearing prices (i.e., demand bids with strictly higher price compared to the MCP must be fully accepted, demand bids with strictly lower price compared to the MCP must be fully rejected, supply bids with strictly lower price compared to the MCP must be fully accepted, supply bids with strictly higher price compared to the MCP must be fully rejected).
We assume that if the MCP is not unique but bounded both from above and below, the resulting clearing price will be equal to the average of the two bounds.
%Furthermore, if the MCP is not unique, but bounded from one side, we assume that the MCP is equal to the bound.

The total social welfare (TSW) is interpreted as the total utility of consumption minus the total cost of production, summed for all products, as described by eq. (\ref{eq:TSW}).

\begin{equation}
    TSW=\sum_{B_E,B_{R+},B_{R-}}x_i q_i p_i
\label{eq:TSW}
\end{equation}

$B_E$: set of energy bids, $B_{R+}$ set of positive reserve bids, $B_{R-}$ set of negative reserve bids. The TSW may be decomposed into bid surplus and congestion rent.

\section{Results}

An illustrative example is presented to highlight the counterintuitive implications of reserve overprocurement.

\subsection{Network topology, parameters and bid sets}

\begin{figure}[h!]
    \centering
    \includegraphics[width=0.4\linewidth]{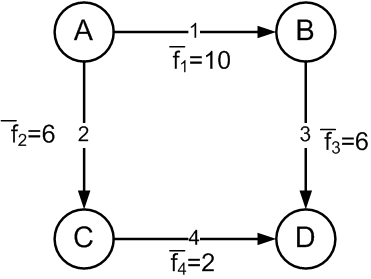}
    \caption{The topology of the example network. Reference directions for positive sign flows are denoted with arrows. Maximal flow values of lines are indicated by $\overline{f}_i$, and are assumed to be identical for both the positive and negative direction of the line.}
    \label{fig:4_zones_new_base}
\end{figure}

We consider the network topology depicted in Fig. \ref{fig:4_zones_new_base}, where we assume that all lines have equal admittance and the bid set summarized in Table \ref{tab:bids_1}. We assume that all bids are divisible, i.e., no block orders are present.

We assume equal admittance of the considered lines.
The respective PTDF of the network is described by the matrix described in eq. (\ref{eq:ptdf}), assuming zone A as the reference (slack) zone. The element $(i,j)$ of the matrix describes the load put on line $j$, when transferring 1 unit of energy from the reference node to node $i$ ($i=1$ corresponds to zone B, etc.). Negative values correspond to flows implied in the opposite direction compared to the reference direction of the line.

\begin{equation}
PTDF=\left[ \begin{array}{cccc}
     3/4 & 1/4 &-1/4 & 1/4  \\
     1/4 & 3/4 & 1/4 & -1/4 \\
     1/2 & 1/2 & 1/2 & 1/2 \\
\end{array}
\right]
    \label{eq:ptdf}
\end{equation}

\begin{table}[h!]
    \centering
    \begin{tabular}{|c|c|c|c|}
     \hline
         type & zone & quantity ($q$) [MWh] & price ($p$) [EUR/MWh]\\ \hline
         E & A & -4 & 12\\ 
         E & A & -4 & 14\\ 
         E & B & 8 & 20\\ 
         E & B & -8 & 18\\ 
         R+ & A & -4 & 1\\         
         R+ & B & 4 & 8\\
         R+ & B & -4 & 6\\
         R- & B & 4 & 6\\
         R- & C & -4 & 4 \\
         \hline
    \end{tabular}
    \caption{Bid set of the example. Products: $E$ energy, $R+$ positive reserve, $R-$ negative reserve.}
    \label{tab:bids_1}
\end{table}

%\newpage

\subsection{Decoupled markets}
\label{subsec_decoupled}
Let us first examine the case of decoupled markets. In this case, the total quantity of accepted demand and supply bids for all products (energy, positive and negative reserve) must be equal for each zone, so no network flows arise, independent of the reserve activation pattern.
The resulting market-clearing prices are summarized in Table \ref{tab:MCP_decoupled}, while the resulting bid acceptance indicators and bid surpluses are summarized in Table \ref{tab:bids_acc_decoupled}.

\begin{table}[h!]
    \centering
    \begin{tabular}{|c|c|c|c|}
    \hline
         zone & $MCP_E$ & $MCP_{R+}$  & $MCP_{R-}$ \\ \hline
         A    & $<$12  & 1 &  - \\
         B    & 19  & 7 & $>$6 \\
         C    &  -  & - & $<$4 \\
         D    &  -  & - &  - \\
         \hline
    \end{tabular}
    \caption{Zonal market-clearing prices of products in the decoupled case.}
    \label{tab:MCP_decoupled}
\end{table}

\begin{table}[h!]
    \centering
    \begin{tabular}{|c|c|c|c|c|c|}
     \hline
         type & zone & quantity ($q$) & price ($p$) & acc. indicator ($x$) & surplus \\ \hline
         E & A & -4 & 12  & 0   & 0\\ 
         E & A & -4 & 14  & 0   & 0\\   
         E  & B & 8  & 20 & 1   & 8\\ 
         E  & B & -8 & 18 & 1   & 8\\ 
         R+ & A & -4 & 1  & 0   & 0\\
         R+ & B & 4  & 8  & 1   & 4\\\
         R+ & B & -4 & 6  & 1   & 4\\
         R- & B & 4  & 6  & 0   & 0\\\
         R- & C & -4 & 4  & 0   & 0 \\
         \hline
    \end{tabular}
    \caption{Resulting bid acceptance indicators and bid surpluses in the decoupled case.}
    \label{tab:bids_acc_decoupled}
\end{table}

In the decoupled case, there is no congestion rent, thus the TSW equals to the sum of the bid surpluses, i.e., 24 units (EUR), from which 16 units originate from energy trade, and 8 units originate from the trading of positive reserves.

\subsection{Optimal co-allocation of energy and reserves, with balances for all products}
\label{subsec_coall_ref}

Let us now use the optimal co-allocation algorithm described in \cite{divenyi2024potential} to calculate a dispatch. In this case, the bid acceptances and market clearing prices are determined to optimize the total welfare, while also taking into account the maximal transfer capacity of network elements. The clearing framework proposed in \cite{divenyi2024potential} may be explained via the concept of \emph{worst-case flows}. Worst-case flows are interpreted in the context of individual network elements and correspond to the flows that arise when the reserve activation pattern turns out to be the 'worst' regarding the load of the network element (line) in question, i.e., it implies the most possible load for the given line. Worst case flows can be interpreted both in the positive and in the negative direction of the line.
The algorithm gives priority to those inter-zonal trades, which induce the most welfare increase, while limiting the flows and potential flows to respect network transmission constraints. The potential (per unit) benefit of inter-zonal trades may be easily derived based on the difference of bid prices. Accordingly, transporting the first 4 units of energy from A to B implies a 6 EUR/MWh benefit compared to the decoupled case, since without interzonal trading, the supply price is 18 EUR/MWh, while A has a source of 4 MWh for 12 EUR/MWh.
Similarly, regarding the second energy bid in zone A, at the price of 14, transporting the second 4 units of energy from A to B implies a 4 EUR/MWh benefit.
Transporting 1 unit of positive reserve from A to B implies a 5 EUR/MWh benefit (since it reduces the original cost of 6 to 1).

\begin{figure}[h!]
    \centering
    \includegraphics[width=0.8\linewidth]{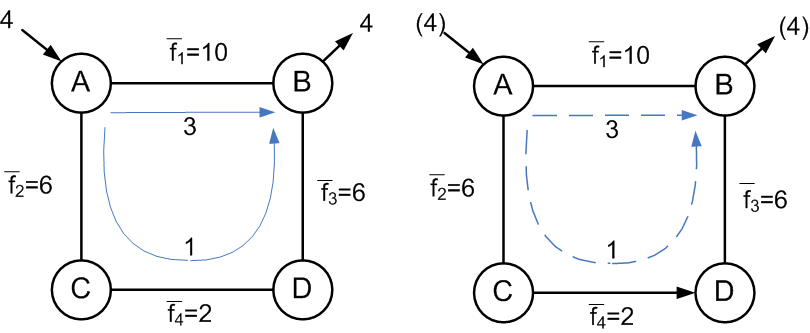}
    \caption{The flows implied in the network, by the cross-zonal allocation of 4 units of energy (left) or 4 units of positive reserve  (right) from zone A to zone B. Numbers in parentheses refer to inlets in the case of reserve activation. Arrows pointing inside of the circles representing the zones indicate inlets corresponding to energy supply or reserve supply activation, while arrows pointing outward correspond to energy consumption and arising reserve demand.    
    Dashed flows correspond to reserve-activation related flows, arising in the case of full activation of the reserves.}
    \label{fig_4_zones_new_flow_4EAB}
\end{figure}

According to the PTDF described in eq. (TBD), transporting 4 units of energy or 4 units of positive reserve implies the same flows in the network (the cross-zonal reserve allocation implies flows, of course, only if it's activated), depicted in Fig. \ref{fig_4_zones_new_flow_4EAB}.

\begin{figure}[h!]
    \centering
    \includegraphics[width=0.4\linewidth]{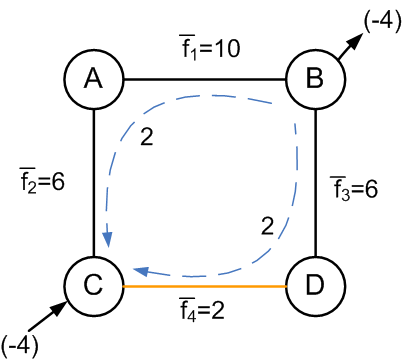}
    \caption{The flows of the network, implied by the cross-zonal allocation of 4 units of negative reserve from zone C to zone B in the case of full activation. Orange color indicates a line load on the capacity limit.}
    \label{fig_4_zones_new_flow_4NRCB}
\end{figure}

In addition, a transport of 4 units of negative reserve from C to B is allowed.
If activated, this reserve implies the flows depicted in Fig. \ref{fig_4_zones_new_flow_4NRCB}.

We can see that the bottleneck is line 4, regarding the flows in the positive direction. Energy and positive reserve related flows discussed above potentially put a load of 1 unit on this line in the positive direction, while the activation of the negative reserve puts load on this line in the negative direction (which would alleviate the congestion by netting). Accordingly, the worst-case load of this line corresponds to the scenario, when positive reserve is maximally activated in node B, and negative reserve is not activated.

The maximal capacity of the line is only 2 units. Accordingly, only 2 of the potential 3 (4+4 units of energy, and 4 units of positive reserve) transfers, which use the line in the positive direction may be allowed. These transfers will yield the highest benefit, i.e., 4 units of energy and 4 units of positive reserve will be transported from zone A to zone B.

\begin{figure}[h!]
    \centering
    \includegraphics[width=0.4\linewidth]{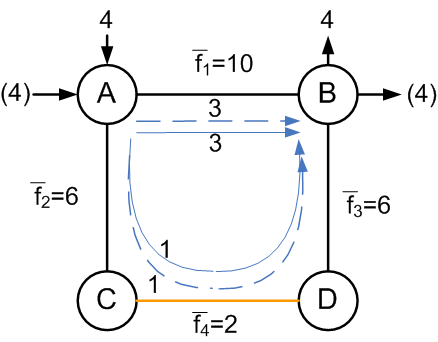}
    \caption{The worst-case flows in the network, in the case of transporting  4 units of energy (continuous line) and 4 units of positive reserve from zone A to zone B (dashed line). Line 4 at its limit load is indicated with orange color.}
    \label{fig_4_zones_new_WCF_4EAB}
\end{figure}

The resulting worst-case flows are depicted in Fig. \ref{fig_4_zones_new_WCF_4EAB},

The resulting market-clearing prices are summarized in Table \ref{tab:MCP_ref}, while the resulting bid acceptance indicators and bid surpluses are summarized in Table \ref{tab:bids_acc_ref}.

\begin{table}[h!]
    \centering
    \begin{tabular}{|c|c|c|c|}
    \hline
         zone & $MCP_E$ & $MCP_{R+}$  & $MCP_{R-}$ \\ \hline
         A    & 12  &  1 &  - \\
         B    & 18  &  8 &  6 \\
         C    &  -  &  - &  4 \\
         D    &  -  &  - &  - \\
         \hline
    \end{tabular}
    \caption{Zonal market-clearing prices of products in the reference co-allocation case.}
    \label{tab:MCP_ref}
\end{table}

\begin{table}[h!]
    \centering
    \begin{tabular}{|c|c|c|c|c|c|}
     \hline
         type & zone & quantity ($q$) & price ($p$) & acc. indicator ($x$) & surplus \\ \hline
         E & A & -4 & 12  & 1   & 0\\ 
         E & A & -4 & 14  & 0   & 0\\   
         E  & B & 8  & 20 & 1   & 16\\ 
         E  & B & -8 & 18 & 0.5 & 0\\ 
         R+ & A & -4 & 1  & 1   & 0\\
         R+ & B & 4  & 8  & 1   & 0\\\
         R+ & B & -4 & 6  & 0   & 0\\
         R- & B & 4  & 6  & 1   & 0\\\
         R- & C & -4 & 4  & 1   & 0 \\
         \hline
    \end{tabular}
    \caption{Resulting bid acceptance indicators and bid surpluses in the  reference co-allocation case.}
    \label{tab:bids_acc_ref}
\end{table}

In this case, the implied TSW of energy trade is 40 EUR, and the implied TSW values of positive and negative reserve trading are 28 and 8 units, respectively,
resulting in an overall value of 76 EUR. From these 76 units, 16 units are realized as bid surplus (of energy bids) and 60 units are realized as congestion rent.

\subsection{An alternative solution with overprocurement of reserves}
\label{subsec_OP}

Let us now consider the dispatch, the market-clearing prices of which are summarized in Table \ref{tab:MCP_OP}, while the corresponding bid acceptance indicators and bid surpluses are summarized in Table \ref{tab:bids_acc_OP}.

\begin{table}[h!]
    \centering
    \begin{tabular}{|c|c|c|c|}
    \hline
         zone & $MCP_E$ & $MCP_{R+}$  & $MCP_{R-}$ \\ \hline
         A    & 14  &  1 &  - \\
         B    & 18  &  6 &  6 \\
         C    &  -  &  - &  4 \\
         D    &  -  &  - &  - \\
         \hline
    \end{tabular}
    \caption{Zonal market-clearing prices of products in the reference co-allocation case.}
    \label{tab:MCP_OP}
\end{table}

\begin{table}[h!]
    \centering
    \begin{tabular}{|c|c|c|c|c|c|}
     \hline
         type & zone & quantity ($q$) & price ($p$) & acc. indicator ($x$) & surplus \\ \hline
         E & A & -4 & 12  & 1   & 8\\ 
         E & A & -4 & 14  & 1   & 0\\   
         E  & B & 8  & 20 & 1   & 16\\ 
         E  & B & -8 & 18 & 0 & 0\\ 
         R+ & A & -4 & 1  & 1   & 8\\
         R+ & B & 4  & 8  & 1   & 0\\\
         R+ & B & -4 & 6  & 0.5   & 0\\
         R- & B & 4  & 6  & 1   & 0\\\
         R- & C & -4 & 4  & 1   & 0 \\
         \hline
    \end{tabular}
    \caption{Resulting bid acceptance indicators and bid surpluses in the  reference co-allocation case.}
    \label{tab:bids_acc_OP}
\end{table}

One may note that, according to this dispatch configuration, all the critical transfers are allowed. In addition to the energy supply bid of 4 units at the price of 12 EUR, 4 additional units are also dispatched in B at the price of 14 EUR. Accordingly altogether, 8 units of energy and 4 units of positive reserve are transferred from zone A to zone B. The flows indicated by these transfers in the case of the activation of this reserve would put a load of 3 on line 4 in the positive direction (while its maximal capacity is 2), which seemingly violates the network transmission constraints.

\begin{figure}[h!]
    \centering
    \includegraphics[width=0.4\linewidth]{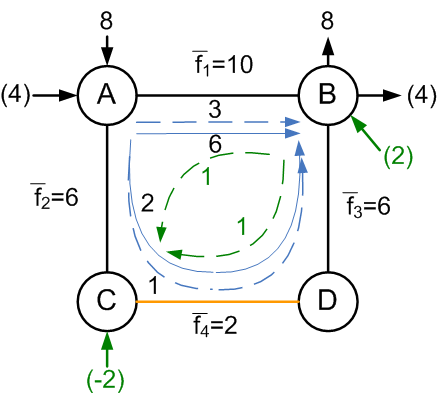}
    \caption{The worst-case flows in the network, in the case of transporting  8 units of energy (continuous line) and 4 units of positive reserve from zone A to zone B (dashed line). Negative reserve supply allocated in node C and positive reserve supply allocated in B are used for congestion management, denoted with green color.}
    \label{fig_4_zones_new_WCF_OP}
\end{figure}

Let us however, consider the following.
This worst-case scenario assumes that the maximal positive reserve demand arises in zone B, which implies that no negative reserve demand is present. This means that 2 units of negative reserve supply allocated in node C (out of the total 4) may be freely used for purposes other than balancing.
Suppose these 2 units of negative reserve supply in node C is co-activated with the 2 units of positive reserve supply allocated in node B. In that case, the resulting flows alleviate the congestion of line 4 as depicted in Fig. \ref{fig_4_zones_new_WCF_OP}.

In this case, the implied TSW of energy trade is 56 EUR, and the implied TSW values of positive and negative reserve trading are 16 and 8 units, respectively,
resulting in an overall value of 80 EUR. From these 80 units, 32 units are realized as bid surplus (of energy bids) and 48 units are realized as congestion rent.
Let us note that compared to the results described in subsection \ref{subsec_coall_ref}, both the total TSW value and the total bid surplus have increased (80 vs 76, and 16 vs 32, respectively), but the congestion rent is decreased by 12 units.

\newpage
\section{Discussion}

First of all, let us emphasize that the dispatch described in subsection \ref{subsec_OP} is not reserve demand netting. In reserve demand netting, reserve demand bids of opposite sign are netted to reduce the amount of activated reserve supply. In the proposed example, unused reserve supply bids are activated in order to reduce the congestion of a critical line.
In the following, we discuss some aspects of the proposed dispatch.

\subsection{Cash flows}
Let us consider the payments of the 
dispatch described in subsection \ref{subsec_OP}, compared to the reference solution, described in subsection \ref{subsec_coall_ref}.
Tables \ref{tab:CF_ref} and \ref{tab:CF_OP} summarize the cash flows in the case of the reference solution and the overprocurement dispatch, respectively.

\begin{table}[h!]
    \centering
    \begin{tabular}{|c|c|c|c|}
    \hline
            Cashflow & Energy & R+ & R-\\
            \hline
         Zone A & -48 & -4  & \\ \hline
         Zone B & 72 & 32  & 24\\ \hline
         Zone C &  &   & -16\\ \hline
         Total &  24 & 28  & 8\\ \hline
    \end{tabular}
    \caption{Cashflows in the case of reference coallocation, whith product balances, as described in subsection \ref{subsec_coall_ref}.}
    \label{tab:CF_ref}
\end{table}

\begin{table}[h!]
    \centering
    \begin{tabular}{|c|c|c|c|}
    \hline
            Cashflow & Energy & R+ & R-\\
            \hline
         Zone A & -112 & -4  & \\ \hline
         Zone B &  144 & 12  & 24\\ \hline
         Zone C &  &   & -16\\ \hline
         Total &  32 & 8  & 8\\ \hline
    \end{tabular}
    \caption{Cashflows in the case of reference coallocation, whith product balances, as described in subsection \ref{subsec_OP}.}
    \label{tab:CF_OP}
\end{table}

As one may notice, the incomes cover the expenses for each product, also in the case of overprocurement.
The total balance in the payments of the participants corresponds to the congestion fee, which is decreased in the overprocurement scenario (as discussed earlier).

\subsection{Potential effect of activation costs}

As discussed previously, no activation costs have been assumed in the current study. If these costs arise and are significant, their effect on the resulting costs and benefits may not be neglected.
However, activation costs are typically not known at the time of allocation.
If one compares the worst-case scenarios in subsections \ref{subsec_coall_ref} and \ref{subsec_OP}, one may notice that more reserve supply activation is present in the latter case. Worst-case scenarios appear, however, very seldom, thus their effect on the expected activation costs is very limited.

\section{Conclusions}

In this work, we used an illustrative example to show that overprocurement of reserves in the portfolio-bidding framework may have beneficial implications. Analysis of the worst-case scenarios, corresponding to the highest possible line loads, has shown that in such critical cases, the additional reserve supplies present due to the overprocurement may be used for congestion management of lines with the highest load. The key finding is that the value brought by the extra transfers that may be allowed on these lines may overwhelm the cost implied by the allocation of additional reserves, thus increasing the overall welfare of the market (although congestion fees are reduced in this case).

\subsection{Future work}
The example included in the current work illustrates that overprocurement may have the potential to increase the resulting welfare in certain cases. It is, however, non-trivial to recognize the scenarios in which such an approach may be applicable.
To decide whether overprocurement may bring benefits, a complex computational framework must be formulated, which describes that in the case of every possible reserve demand activation pattern (including the 'worst-case' scenarios, which maximize the line loads), the remaining reserve supplies may be used to generate flows, which alleviate congestion of the critical lines.
If activation costs are known in advance, at the time of the reserve allocation, their consideration in the allocation process (i.e., their expected value) will add a further level of complexity to the required computational framework.

\section{Acknowledgements}

This work has been supported 
%by the Hungarian Academy of Sciences under its Momentum Programme LP2021-2 and 
by the Fund FK 137608 of the Hungarian National Research, Development and Innovation Office.

\newpage
\bibliographystyle{plain}
\bibliography{references}

\end{document}